\documentclass[letterpaper, 10 pt, conference]{ieeeconf}  % Comment this line out if you need a4paper

\IEEEoverridecommandlockouts                              % This command is only needed if 
\usepackage{epsfig} % for postscript graphics files
\usepackage{cite}

\title{\LARGE \bf
Changing human's impression of empathy from agent by verbalizing agent's position
}

\author{Takahiro Tsumura$^{1}$ and Seiji Yamada$^{1}$ % <-this % stops a space
\thanks{$^{1}$Takahiro Tsumura and Seiji Yamada is with Department of Informatics, The Graduate University for Advanced Studies, SOKENDAI, Tokyo, Japan.
Also Takahiro Tsumura and Seiji Yamada is with Digital Content and Media Sciences Research Division, National Institute of Informatics, Tokyo, Japan.
        }%
}
\begin{document}
\maketitle
\thispagestyle{empty}
\pagestyle{empty}
%%%%%%%%%%%%%%%%%%%%%%%%%%%%%%%%%%%%%%%%%%%%%%%%%%%%%%%%%%%%%%%%%%%%%%%%%%%%%%%%
\begin{abstract}
As anthropomorphic agents (AI and robots) are increasingly used in society, empathy and trust between people and agents are becoming increasingly important. 
A better understanding of agents by people will help to improve the problems caused by the future use of agents in society. 
In the past, there has been a focus on the importance of self-disclosure and the relationship between agents and humans in their interactions.
In this study, we focused on the attributes of self-disclosure and the relationship between agents and people. 
An experiment was conducted to investigate hypotheses on trust and empathy with agents through six attributes of self-disclosure (opinions and attitudes, hobbies, work, money, personality, and body) and through competitive and cooperative relationships before a robotic agent performs a joint task. 
The experiment consisted of two between-participant factors: six levels of self-disclosure attributes and two levels of relationship with the agent. 
The results showed that the two factors had no effect on trust in the agent, but there was statistical significance for the attribute of self-disclosure regarding a person's empathy toward the agent. 
In addition, statistical significance was found regarding the agent's ability to empathize with a person as perceived by the person only in the case where the type of relationship, competitive or cooperative, was presented.
 The results of this study could lead to an effective method for building relationships with agents, which are increasingly used in society.
\end{abstract}

%%%%%%%%%%%%%%%%%%%%%%%%%%%%%%%%%%%%%%%%%%%%%%%%%%%%%%%%%%%%%%%%%%%%%%%%%%%%%%%%
\section{INTRODUCTION}
Humans live in society and use a variety of tools such as ChatGPT and generative image AI, and today's AI issues are focused on the trustworthy and ethical use of AI technology.
Kaplan et al.~\cite{Kaplan23} aimed to identify the key factors that predict trust in AI, and they examined three predictive categories and subcategories from data from 65 papers: human characteristics and capabilities, AI performance and attributes, and contextual challenges. 
All categories examined were significant predictors of trust in AI. 
The more AI is used in human society, the more trust in AI is discussed, and a key issue is that the failure to establish appropriate trust relationships leads to overconfidence and distrust in AI agents, which in turn leads to poor task performance.

When performing tasks with an AI, it is also important to consider ways to repair trust in the AI. 
Kähkönen et al.~\cite{Kähkönen21} systematically reviewed trust repair research conducted over the past two decades to provide researchers and managers with comprehensive insights and directions for future research. 
The review suggested that early use of trust repair strategies for small violations can prevent these violations from escalating into larger violations and, in turn, increase the efficiency and effectiveness of trust repair with employees. 
Bunting et al.~\cite{Bunting21} used insights from focus group discussions on newly created trust, mistrust, and distrust questions to identify how citizens perceive these different concepts and how these perceptions are gendered. 
They then used the new survey data they collected to examine how the focus group results influenced survey responses and which survey items were most likely to effectively measure the three concepts.

Along with trust, we often empathize with artifacts. 
Humans are known to have a tendency to treat artifacts as if they were human in the media equation. 
However, some humans do not accept these agents~\cite{Nomura16}. 
Empathy is closely related to trust. 
As agents permeate society in the future, it is hoped that they will have an element of human acceptance. 
Omdahl~\cite{Omdahl95} broadly classified empathy into three types: (1) affective empathy, which is an emotional response to the emotional state of another person; (2) cognitive empathy, defined as a cognitive understanding of the emotional state of another person; and (3) empathy that includes the above two types of empathy. 
Preston and de Waal~\cite{Preston02} suggested that at the heart of empathic response is a mechanism that allows the observer to access the subjective emotional state of the subject. 
They defined the Perception Action Model (PAM) and integrated different perspectives in empathy.
They defined empathy as three types: (a) sharing or being influenced by the emotional states of others, (b) evaluating the reasons for emotional states, and (c) having the ability to identify and incorporate other perspectives.

In the present study, we investigated the factors that contribute to trust and empathy toward agents to enhance their evaluation prior to performing a task. 
The definition of empathy in this study is empathy toward an agent, not having the agent acquire empathic competence. 
An agent's empathic capacity in this study is what makes people perceive that the agent can empathize with them. 
Through the self-introductions used in the experiment, participants will be asked to assess whether or not they have empathy for an agent.

\section{RELATED WORK}
We summarized some of the previous studies on trust in AI.
Noting that lack of trust is one of the main obstacles standing in the way of taking full advantage of AI, Gillath et al.~\cite{Gillath21} focused on increasing trust in AI through emotional means. Specifically, they tested the association between attachment style, an individual difference that describes how people feel, think, and act in relationships, and trust in AI. 
Results showed that increasing attachment insecurity decreased trust, while increasing attachment security increased trust in AI. 
Focusing on clinicians as the primary users of AI systems in healthcare, Asan et al.~\cite{Asan20} presented the factors that shape trust between clinicians and AI. 
They focused on key trust-related issues to consider when developing AI systems for clinical use.

With the emphasis on research on trust in human relationships, research on trust in AI agents has also received attention.
In a study of trustworthy AI agents, Maehigashi et al.~\cite{Maehigashi22-1} investigated how beeps emitted by a social robot with anthropomorphic physicality affect human trust in the robot. 
They found that (1) sounds just prior to a higher performance increased trust when the robot performed correctly, and (2) sounds just prior to a lower performance decreased trust to a greater extent when the robot performed inaccurately. 
To determine how anthropomorphic physicality affects human trust in agents, Maehigashi et al.~\cite{Maehigashi22-2} also investigated whether human trust in social robots with anthropomorphic physicality is similar to trust in AI agents and humans. 
Also, they investigated whether trust in social robots is similar to trust in AI agents and humans. 
The results showed that trust in social robots was basically not similar to trust in AI agents or humans, and was entrenched between the two.

Maehigashi~\cite{Maehigashi22-3} experimentally investigated the nature of human trust in communication robots compared to trust in other people and AI systems. 
Results showed that trust in robots in computational tasks that yield a single solution is essentially similar to that in AI systems, and partially similar to trust in others in emotion recognition tasks that allow multiple interpretations.
Okamura and Yamada~\cite{Okamura20-1} proposed a method for adaptive trust calibration that consists of a framework for detecting inappropriate calibration conditions by monitoring the user's trust behavior and cognitive cues, called ``trust calibration cues," that prompt the user to resume trust calibration. 
Okamura and Yamada~\cite{Okamura20-2} focused their research on trust alignment as a way to detect and mitigate inappropriate trust alignment, and they addressed these research questions using a behavior-based approach to understanding calibration status. 
The results demonstrate that adaptive presentation of trust calibration cues can facilitate trust adjustments more effectively than traditional system transparency approaches.

We also considered the design of empathy factors from previous studies of anthropomorphic agents using empathy. 
Tsumura and Yamada~\cite{Tsumura23-1} focused on self-disclosure from agents to humans in order to enhance human empathy toward anthropomorphic agents, and they experimentally investigated the potential for self-disclosure by agents to promote human empathy. 
Tsumura and Yamada~\cite{Tsumura23-2} also focused on tasks in which humans and agents engage in a variety of interactions, and they investigated the properties of agents that have a significant impact on human empathy toward them. 
To clarify the empathy between agents/robots and humans, Paiva represented the empathy and behavior of empathetic agents (called empathy agents in HAI and HRI studies) in two different ways: targeting empathy and empathizing with observers~\cite{Paiva17}.
Rahmanti et al.~\cite{Rahmanti22} designed a chatbot with artificial empathic motivational support for dieting called ``SlimMe'' and investigated how people responded to the diet bot.
They proposed a text-based emotional analysis that simulates artificial empathic responses to enable the bot to recognize users' emotions.
Jáuregui et al.~\cite{Jáuregui21} found evidence to support or oppose a robotic mirroring framework, resulting in increased interest in self-tracking technology for health care. 

We also present some previous studies on the attributes of self-disclosure and its relationship to agents.
Pan et al.\cite{Pan20} examined the effect of exposure to online support-seeking posts containing different levels of self-disclosure depth (baseline, peripheral, core) affecting the quality (person-centeredness and politeness) of participants' messages providing support.
A study of cooperative and competitive tasks was conducted by Ruissen and de Bruijn~\cite{Ruissen16}. 
In the study, cooperative and competitive tasks were tested using Tetris. 
The results confirmed that the cooperative task did not reduce self-integration, but the competitive task did. 

Collins and Miller~\cite{Collins94} aimed to clarify and review this literature by using a meta-analytic procedure to evaluate the evidence for three different disclosure preference effects.
For each effect, they found a significant relationship favoring disclosure.
(a) people who make intimate disclosures tend to be liked more than those who disclose at lower levels; (b) people disclose more to those they initially favor; and (c) people like other people as a result of disclosing information to them.
Meng and Dai~\cite{Meng21} investigated when and how effective chatbot emotional support is in reducing people's stress and worry.
The results showed that the positive effects of emotional support on anxiety reduction were enhanced when conversation partners mutually self-disclosed.

\section{MATERIALS AND METHODS}
\subsection{Hypotheses}
The purpose of this study is to examine whether self-disclosure attributes and types of relationships with participants influence trust and empathy toward agents, focusing on the time before agents perform tasks with participants. 
This objective could be an important factor in promoting the use of agents in society. 
The following hypotheses are formulated in this study. 
If these hypotheses are supported, this research will be valuable in developing agents that are more acceptable to humans. On the basis of the above, four hypotheses were examined.

\begin{enumerate}
\item[\textbf{H1}:] Trust toward the agent changes with the attribute of self-disclosure.
\item[\textbf{H2}:] Trust toward the agent increases when a cooperative relationship is presented rather than a competitive relationship.
\item[\textbf{H3}:] Empathy toward the agent changes with the attribute of self-disclosure, but does not change with the type of relationship presented.
\item[\textbf{H4}:] The empathy capacity of the agent does not change with the attribute of self-disclosure, but it changes with the type of relationship presented.
\end{enumerate}

H1 and H2 investigate whether trust in an agent is affected by the two factors in this study, that is, six levels of self-disclosure attributes and two levels of relationship with the agent\cite{Collins94,Meng21}. 
Previous studies with people have shown that the content of self-disclosure affects trust, and that cooperative relationships are more likely to gain trust than competitive relationships. 
For this reason, we investigate whether trust in agents changes similarly to that in people. 

Previous studies by Tsumura and Yamada~\cite{Tsumura23-1,Tsumura23-2} have shown that H3 is a key factor in the relationship between trust and self-disclosure, and the present study will investigate the impact of each factor in detail. 
H4 is an investigation of agents' capacity for empathy. 
Previous studies have shown that the amount of self-disclosure affects empathy, and since the amount of self-disclosure in this study is almost the same, we assumed that it does not affect the ability to empathize~\cite{Pan20}. 
Regarding the type of relationship presented, we also considered that the participant's evaluation of the agent's empathy ability would change as the impression of the agent changes depending on the relationship~\cite{Ruissen16}.

\subsection{Experimental details}
\begin{figure*}[tpb]
\centering
    \includegraphics[scale=0.45]{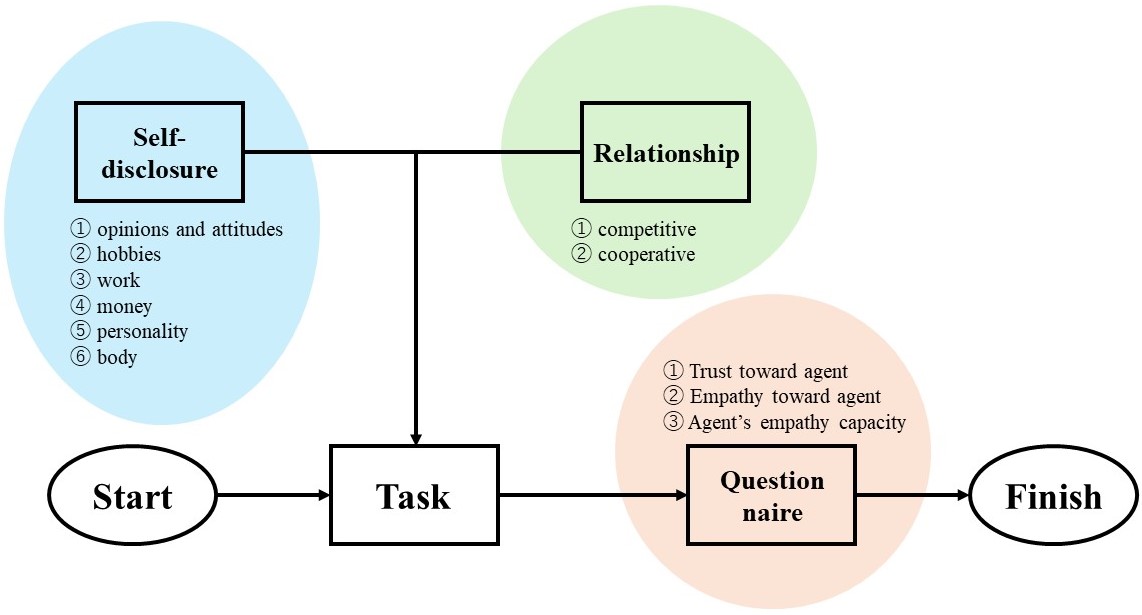}
    \caption{Flowchart of experiment.}
    \label{fig1}
\end{figure*}

The experiment consisted of a video viewing task, followed by three different surveys. 
The experiment was conducted in an online environment using Yahoo! Crowdsourcing. 
The online environment used in this experiment has already been used as one experimental environment~\cite{Davis99,Tsumura23-1,Tsumura23-2}. 
A flowchart of this experiment is shown in Figure ~\ref{fig1}. 
Participants learned about self-disclosure attributes and their relationship with the agent by watching a video of the agent giving a self-introduction. 
This study's agent was a learning support agent, and the situation is that the agent is giving a self-introduction before providing learning support. 
In the videos, the agent used the same gestures regardless of the factors, and the videos were silent, so self-introductions were done via text message. 
None of the videos were longer than 2 minutes, and participants could not proceed to a questionnaire without confirming that they had watched the video.

Because the experiment was conducted on a participant-to-participant factorial basis, all participants watched only one video out of 12 combinations and completed the questionnaire. 
The questionnaire consisted of three types of questions: trust in the agent, empathy toward the agent, and the agent's ability to empathize. 
In addition, one question involved a self-disclosure check used in the task and a confirmation that the participant had watched the video.

The experiment was conducted at 12 levels in total, with 6 levels of self-disclosure attributes and 2 levels of relationship with the agent as 2 factors. 
The independent variables were self-disclosure attributes (opinions and attitudes, hobbies, work, money, personality, and body) and relationship with the agent (competitive and cooperative). 
The three dependent variables were trust in the agent, empathy toward the agent, and the agent's capacity for empathy.

\subsection{Participants}
We recruited participants through Yahoo! Crowdsourcing and paid 40 yen (=0.27 dollars) per participant as compensation. 
We created a web page for the experiment using Google Forms and uploaded and embedded a video created for the experiment on YouTube.

There was a total of 587 participants. 
Cronbach's $\alpha$ coefficient was then used to determine the reliability of the trust questionnaire, and the coefficient was determined to be 0.9160-0.9612 in all conditions. 
In addition, Cronbach's $\alpha$ coefficient was used for the reliability of the empathy questionnaire for agents, and the coefficients were determined to be 0.6965 to 0.8224. 
Cronbach's $\alpha$ coefficient was used for the reliability of the questionnaire for empathy capacity of agents, and the coefficients were determined to range from 0.7904 to 0.8680.

In the analysis, 47 participants in each of all conditions were analyzed in the order of their participation. 
Thus, the total number of participants used in the analysis was 564. 
The mean age was 48.90 years (standard deviation = 11.03 years), with a minimum age of 20 years and a maximum age of 79 years. There were also 432 males and 132 females.

\subsection{Self-disclosure attributes}
In this study, we prepared six levels of self-disclosure attributes. 
The reason for preparing six levels was to investigate the impression that the agent's self-disclosure gives to others, on the basis of previous studies that have classified self-disclosure into six categories~\cite{Jourard71}. 
The content of each type of self-disclosure is as follows. 
In the content related to opinions and attitudes, the agent talked about the most important factors in their learning. 
For hobbies, the agent talked about its love of learning. 
For work, the agent talked about the pressures of work in learning support. 
For money, the agent discussed how money is spent on learning.
 For personality, the agent talked about what people say about its personality.
 For body, the agent talked about its own appearance.

In all conditions, the amount of conversation was kept nearly uniform, and participants spent approximately the same amount of time watching the videos. 
Each attribute of self-disclosure was recognized by the participants; the results of the chi-square test showed that $\chi^2$(25)=1043, $p<0.01$, and the results of the residual analysis showed that the participants correctly recognized the attribute of self-disclosure. 
On the basis of these results, a manipulation check for each attribute of self-disclosure was completed, and this experimental data was used for analysis.

\subsection{Relationship with agent} 
\begin{figure}[tbp]
		\begin{center}
		\includegraphics[width=76mm]{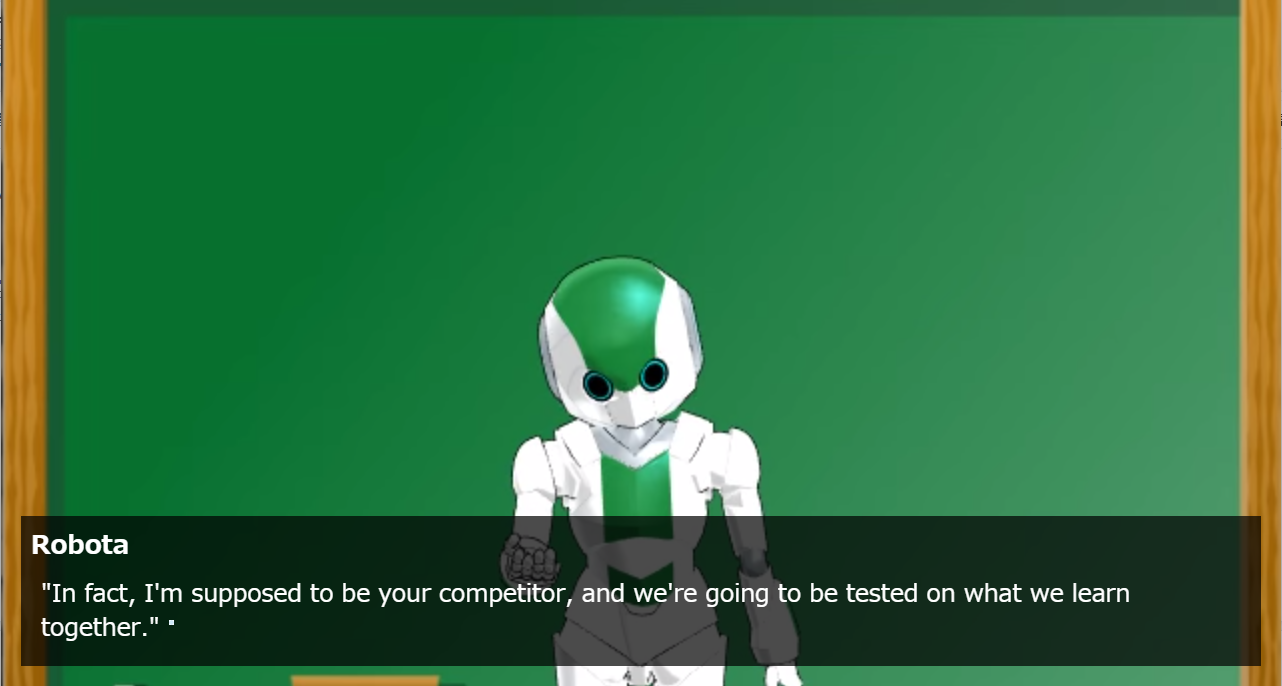}
		\caption{Competitive agent}
		\label{fig2}
	\end{center}
\end{figure}

\begin{figure}[tbp]
		\begin{center}
		\includegraphics[width=76mm]{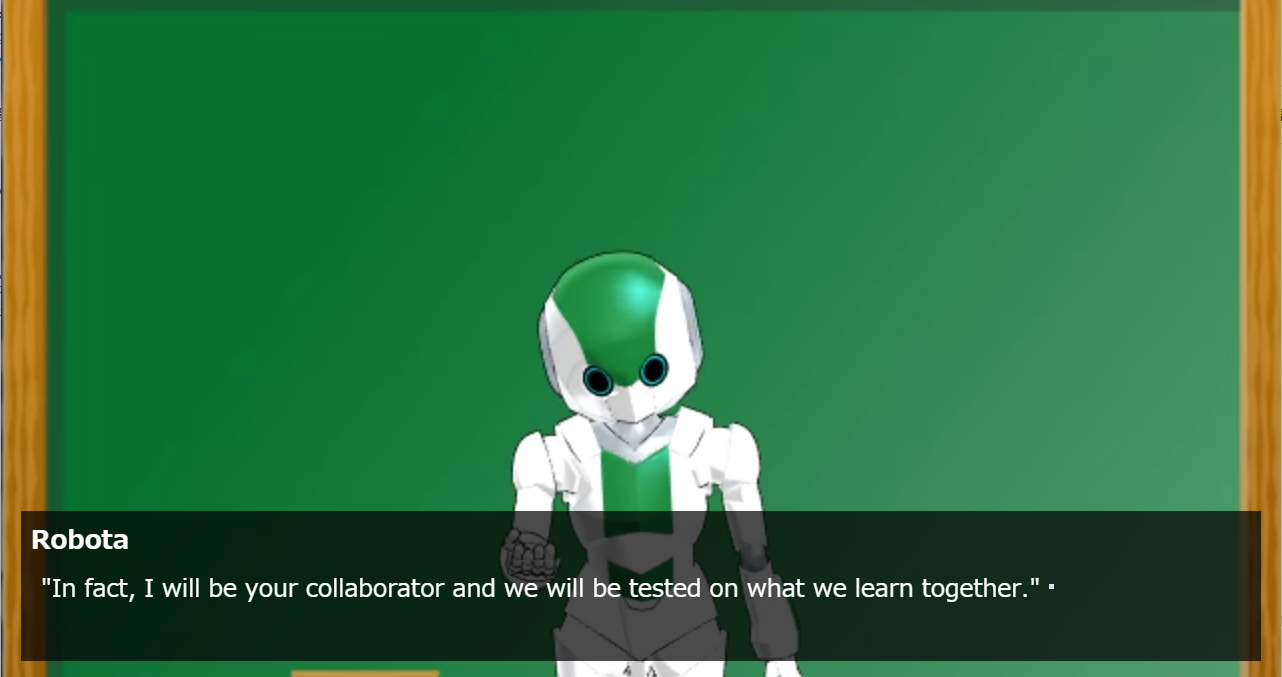}
		\caption{Cooperative agent}
		\label{fig3}
	\end{center}
\end{figure}

There were two levels of relationship with the agent. 
The agent either conveyed that it was a competitor or that it was a cooperative partner. 
The purpose of this experiment was not to actually perform a joint task but to investigate the change in the participant's impression of the agent by only presenting the type of relationship. 
The competing agent is shown in Figure ~\ref{fig2}, and the cooperating agent is shown in Figure ~\ref{fig3}. 
The agents in this task were introduced as learning support agents, but the relationship is mentioned at the end of their self-introduction. 
For this reason, the content of each attribute of self-disclosure and the topic of the relationship with the agent are independent.

\subsection{Questionnaire}
In this study, we used a questionnaire related to empathy that has been used in previous psychological studies. 
To measure cognitive trust, the Multi-Dimensional Measure of Trust (MDMT)~\cite{Ullman19} was used. MDMT was developed to measure a task partner's reliability and competence corresponding to the definition of cognitive trust.
The participants rated how much the partner AI fit each word (reliable, predictable, dependable, consistent, competent, skilled, capable, and meticulous) on an 8-point scale (0: not at all - 7: very). 
Moreover, for emotional trust, we asked participants to answer how much the partner AI fit each word (secure, comfortable, and content) on a 7-point scale (1: strongly disagree - 7: strongly agree) as in the previous study~\cite{Komiak06}.
In our study, we removed the matching 0 scale of cognitive trust, bringing it to the same 7 scale as emotional trust.
The trust questionnaire used in this study was the one used by Maehigashi et al.~\cite{Maehigashi22-1}.

To investigate the characteristics of empathy, we modified the Interpersonal Reactivity Index (IRI) to be an index for anthropomorphic agents. 
The main modifications were changing the target name. 
In addition, the number of items on the IRI was modified to 12; for this, items that were not appropriate for the experiment were deleted, and similar items were integrated. 
Since both of the questionnaires used were based on IRI, a survey was conducted using a 5-point Likert scale (1: not applicable, 5: applicable). 

The questionnaire used is shown in Table~\ref{table1}. 
Since Q4, Q9, and Q10 were reversal items, the points were reversed during analysis. Q1 to Q6 were related to affective empathy, and Q7 to Q12 were related to cognitive empathy. 
Participants answered a questionnaire after completing the task. Qet is a questionnaire on empathy for the agent, and Qeo is a questionnaire on the agent's capacity for empathy.

\renewcommand{\arraystretch}{1.0}
\begin{table*}[tbp] 
    \caption{Summary of questionnaire used in this experiment}
    \centering
    \scalebox{0.9}{
    \begin{tabular}{llll}\hline 
        \multicolumn{4}{l}{\textbf{Trust}}\\ \hline
        \multicolumn{4}{l}{\textbf{Cognitive trust}}\\
        Qt1: Reliable. & Qt2: Predictable. & Qt3: Dependable. & Qt4: Consistent. \\
        Qt5: Competent. & Qt6: Skilled. & Qt7: Capable. & Qt8: Meticulous. \\
        \multicolumn{4}{l}{\textbf{Emotional trust}}\\
        Qt9: Secure. & Qt10: Comfortable. & Qt11: Content. & \\\hline
        \multicolumn{4}{l}{\textbf{Affective empathy}}\\ \hline
        \multicolumn{4}{l}{\textbf{Personal distress}}\\
        \multicolumn{4}{l}{Qet1: If an emergency happens to the robot, you would be anxious and restless.}\\
        \multicolumn{4}{l}{Qet2: If the robot is emotionally disturbed, you would not know what to do.}\\
        \multicolumn{4}{l}{Qet3: If you see the robot in need of immediate help, you would be confused and would not know what to do.}\\
        \textbf{Empathic concern}\\
        \multicolumn{4}{l}{Qet4: If you see the robot in trouble, you would not feel sorry for that robot.}\\
        \multicolumn{4}{l}{Qet5: If you see the robot being taken advantage of by others, you would feel like you want to protect that robot.}\\
        \multicolumn{4}{l}{Qet6: The robot's story and the events that have taken place move you strongly.}\\\hline
        \multicolumn{4}{l}{\textbf{Cognitive empathy}}\\ \hline
        \multicolumn{4}{l}{\textbf{Perspective taking}}\\
        \multicolumn{4}{l}{Qet7: You look at both the robot's position and the human position.}\\
        \multicolumn{4}{l}{Qet8: If you were trying to get to know the robot better, you would imagine how that robot sees things.}\\
        \multicolumn{4}{l}{Qet9: When you think you're right, you don't listen to what the robot has to say.}\\
        \multicolumn{4}{l}{\textbf{Fantasy scale}}\\
        \multicolumn{4}{l}{Qet10: You are objective without being drawn into the robot's story or the events taken place.}\\
        \multicolumn{4}{l}{Qet11: You imagine how you would feel if the events that happened to the robot happened to you.}\\
        \multicolumn{4}{l}{Qet12: You get deep into the feelings of the robot.}\\\hline
        \multicolumn{4}{l}{\textbf{Affective empathy}}\\ \hline
        \multicolumn{4}{l}{\textbf{Personal distress}}\\
        \multicolumn{4}{l}{Qeo1: Do you think the robot would be anxious and restless if an emergency situation happened to you?}\\
        \multicolumn{4}{l}{Qeo2: Do you think the robot would not know what to do in a situation where you are emotionally involved?}\\
        \multicolumn{4}{l}{Qeo3: Do you think the robot will be confused and not know what to do when it sees itself in imminent need of help?}\\
        \multicolumn{4}{l}{\textbf{Empathic concern}}\\
        \multicolumn{4}{l}{Qeo4: Do you think the robot would not feel sorry for you if it saw you in trouble?}\\
        \multicolumn{4}{l}{Qeo5: Do you think the robot would feel like protecting you if it saw you being used by others for their own good?}\\
        \multicolumn{4}{l}{Qeo6: Do you think the robot is strongly moved by your story and the events that took place?}\\\hline
        \multicolumn{4}{l}{\textbf{Cognitive empathy}}\\ \hline
        \multicolumn{4}{l}{\textbf{Perspective taking}}\\
        \multicolumn{4}{l}{Qeo7: Do you think the robot will look at both your position and the robot's position?}\\
        \multicolumn{4}{l}{Qeo8: Do you think the robot tried to get to know you better and imagined how things looked from your point of view?}\\
        \multicolumn{4}{l}{Qeo9: Do you think robots won't listen to your arguments when you seem to be right?}\\
        \multicolumn{4}{l}{\textbf{Fantasy scale}}\\
        \multicolumn{4}{l}{Qeo10: Do you think the robot is objective without being drawn into your story or the events that took place?}\\
        \multicolumn{4}{l}{Qeo11: Do you think robots imagine how they would feel if the events that happened to you happened to them?}\\
        \multicolumn{4}{l}{Qeo12: Do you think the robot will go deeper into your feelings?}\\\hline
    \end{tabular}}
    \label{table1}
\end{table*}

\subsection{Analysis method}
A two-factor analysis of variance was used. ANOVA has been frequently used in previous studies and is the appropriate method of analysis with respect to this study. 
As mentioned above, the factors among participants were six levels of self-disclosure attributes and two levels of relationship with the agent.
On the basis of the results of the participants' questionnaires, we examined how the self-disclosure attributes and relationship with the agent affected trust and empathy toward the agent and the agent's ability to empathize with the participant. 
The trust and empathy values tabulated in the task were used as dependent variables. 
The statistical software R (ver. 4.1.0) was used for the ANOVA and multiple comparisons in all analyses in this paper.

\section{RESULTS}
\begin{table*}[tbp]
 \caption{Results of participants' statistical information}
 \centering
 \scalebox{1.0}{
 \begin{tabular}{c|c|cc||c|cc}\hline 
 Category & Conditions & Mean & S.D. & Conditions & Mean & S.D. \\ \hline 
  & opinions and attitudes-competitive & 50.83 & 10.84 & money-competitive & 48.26 & 9.863 \\ 
  & opinions and attitudes-cooperative & 48.55 & 9.174 & money-cooperative & 48.83 & 10.75\\ 
 Trust& hobbies-competitive & 46.94 & 10.40 & personality-competitive & 51.40 & 10.59 \\ 
(Qt1-Qt11) & hobbies-cooperative & 50.02 & 10.71 & personality-cooperative & 50.60 & 8.561 \\
  & work-competitive & 50.96 & 9.297 & body-competitive & 49.55 & 8.091 \\ 
  & work-cooperative & 49.36 & 11.00 & body-cooperative & 50.85 & 8.718 \\ \hline
  & opinions and attitudes-competitive & 33.96 & 6.406 & money-competitive & 31.62 & 6.152 \\ 
  & opinions and attitudes-cooperative & 32.98 & 6.476 & money-cooperative & 31.66 & 5.913\\ 
 Empathy & hobbies-competitive & 32.32 & 5.146 & personality-competitive & 31.79 & 6.075 \\ 
(Qet1-Qet12) & hobbies-cooperative & 34.02 & 6.229 & personality-cooperative & 32.26 & 6.635 \\
  & work-competitive & 35.43 & 6.590 & body-competitive & 31.70 & 6.947 \\ 
  & work-cooperative & 33.13 & 6.762 & body-cooperative & 32.43 & 6.382 \\ \hline
  & opinions and attitudes-competitive & 35.47 & 7.244 & money-competitive & 32.45 & 6.286 \\ 
  & opinions and attitudes-cooperative & 33.81 & 8.061 & money-cooperative & 34.60 & 6.765\\ 
 Empathy & hobbies-competitive & 34.57 & 5.763 & personality-competitive & 35.11 & 7.690 \\ 
(Qeo1-Qeo12) & hobbies-cooperative & 37.55 & 6.477 & personality-cooperative & 35.09 & 7.208 \\
  & work-competitive & 33.57 & 6.341 & body-competitive & 32.43 & 6.971 \\ 
  & work-cooperative & 35.64 & 7.394 & body-cooperative & 34.81 & 6.412 \\ \hline
 \end{tabular}} \\ 
 \label{table2}
\end{table*}

\begin{table*}[tbp]
\caption{Analysis results of ANOVA}
\centering
\scalebox{1.0}{
\begin{tabular}{c|llll}\hline
& \multicolumn{1}{c}{Factor} & \multicolumn{1}{c}{\em{F}} & \multicolumn{1}{c}{\em{p}} & \multicolumn{1}{c}{$\eta^2_p$}\\ \hline
& Self-disclosure attributes & 0.9593 & 0.4422 \em{ns} & 0.0086 \\ 
Trust & Relationship with agent & 0.0031 & 0.9558 \em{ns} & 0.0000\\
(Qt1-Qt11)& Self-disclosure attributes $\times$ Relationship with agent & 0.9581 & 0.4430 \em{ns} & 0.0086 \\ \hline
& Self-disclosure attributes & 2.470 & 0.0315 * & 0.0219 \\ 
Empathy & Relationship with agent & 0.0113 & 0.9152 \em{ns} & 0.0000\\
(Qet1-Qet12)& Self-disclosure attributes $\times$ Relationship with agent & 1.158 & 0.3286 \em{ns} & 0.0104 \\ \hline
& Self-disclosure attributes & 1.778 & 0.1156 \em{ns} & 0.0158 \\ 
Empathy & Relationship with agent & 5.106 & 0.0242 * & 0.0092\\
(Qeo1-Qeo12)& Self-disclosure attributes $\times$ Relationship with agent & 1.554 & 0.1715 \em{ns} & 0.0139 \\ \hline
\multicolumn{5}{c}{}
\end{tabular}} \\
\hspace{-110mm}
\em{p}:
{{*}p\textless\em{0.05}}
\label{table3}
\end{table*}

\begin{table*}[tbp]
 \caption{Multiple comparisons of self-disclosure attributes: Mean and S.D.}
 \centering
 \scalebox{1.0}{
 \begin{tabular}{c|c|cc}\hline 
 Category & Conditions & Mean & S.D. \\ \hline 
  & opinions and attitudes & 33.47 & 6.425 \\ 
  & hobbies & 33.17 & 5.747 \\ 
 Empathy & work & 34.28 & 6.740 \\ 
(Qet1-Qet12) & money & 31.64 & 6.002 \\
  & personality & 32.02 & 6.331 \\ 
  & body & 32.06 & 6.644 \\ \hline
 \end{tabular}} \\ 
 \label{table4}
\end{table*}

\begin{figure*}[tbp]
\centering
\includegraphics[scale=0.5]{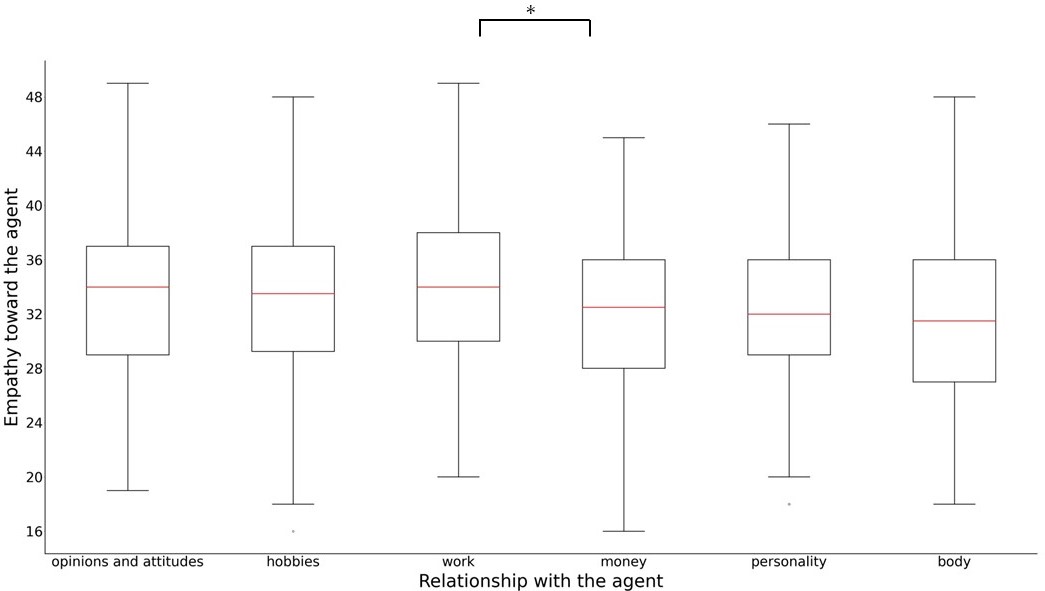}
\caption{Results of multiple comparisons on self-disclosure attributes. Red lines are medians, and circles are outliers.}
\label{fig4}
\end{figure*}

\begin{figure}[tbp]
\centering
\includegraphics[scale=0.3]{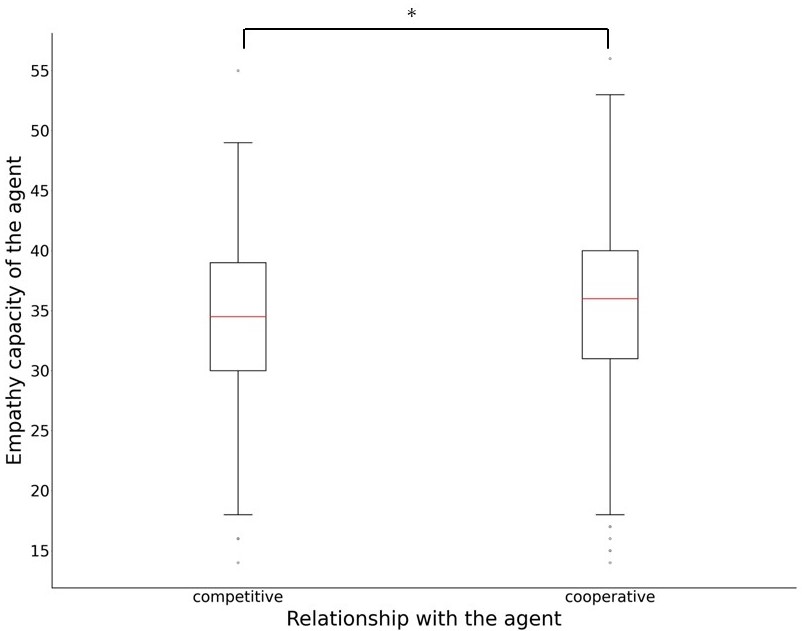}
\caption{Results of main effects on relationships with agent. Red lines are medians, and circles are outliers.}
\label{fig5}
\end{figure}

In this study, cognitive and emotional trust were considered together as trust. 
Table~\ref{table2} shows the mean and standard deviation for each condition. 
Table~\ref{table3} shows the results of the analysis of variance for the three questionnaires. 
In this paper, when a main effect was found, multiple comparisons were examined for significant differences using multiple comparison tests with Tukey's HSD test.

The results showed that trust (Qt1-Qt11) toward the agent was not significant. 
For empathy toward the agent (Qt1-Qt12), a main effect was found for the self-disclosure attribute. 
The results of the multiple comparisons are shown in Figure~\ref{fig4}. 
Table~\ref{table4} shows the mean and standard deviations of the results of the multiple comparisons of empathy for the agents. 
Among the attributes of self-disclosure, significance was found between the job and money content. 
This indicates that self-disclosure attributes affected empathy toward the agent in some cases.

In addition, a main effect was found for the agent's empathy capacity (Qeo1-Qeo12) in relationship with agent. 
The results of the main effect are shown in Figure~\ref{fig5}. 
This allows us to imagine that agents are more empathetic than those who just tell people that they are in a cooperative relationship, but not in a competitive relationship. 
It was shown that people's evaluations of the empathy capacity of the agent were variable even before the actual collaborative task.

\section{DISCUSSION}
%\subsection{Hypothesis summary}
The way to properly build a relationship between a human and an AI agent is not only trust and empathy toward the agent, but also proper recognition by the person of the agent's capacity for empathy. 
Trust and empathy toward agents are necessary for agents to be utilized in society. 
If trust and empathy toward agents can be made constant through appropriate approaches, humans and agents can build a trusting relationship.

In this study, we conducted an experiment to investigate the conditions necessary for humans to trust and empathize with agents. 
We focused on the agent's self-disclosure attributes and the relationship with the agent as factors that influence trust and empathy. 
The purpose of this study was to investigate whether the attributes of self-disclosure and the relationship with the agent can control trust and empathy toward the agent in interactions with the agent. 
To this end, four hypotheses were formulated, and the data from the experiment were analyzed.

The results of the experiment showed that the attribute of self-disclosure and the relationship with the agent did not cause statistically significant changes in trust toward the agent. 
H1, ``Trust toward the agent changes with the attribute of self-disclosure" was not supported. 
In addition, H2, ``Trust toward the agent increases when a cooperative relationship is presented rather than a competitive relationship," was not also supported. 
One possible factor that led to this result is that the amount of information in each of the self-disclosures did not change, despite differences in the attributes of self-disclosure. 
Regarding the relationship with the agent, it is also possible that trust in the agent is related to the agent's ability as demonstrated by the agent actually performing the task, and that simply presenting the relationship with the agent in text messages alone, as in this study, would not be sufficient to make a sufficient difference.

On the other hand, H3, ``Empathy toward the agent changes with the attribute of self-disclosure, but does not change with the type of relationship presented," supported the hypothesis. 
The results for H3 were similar to those obtained by Tsumura and Yamada~\cite{Tsumura23-1,Tsumura23-2}, and the same results were obtained in this study.
In addition, since the agents in this study were engaged in conversation for the purpose of supporting learning, it is possible that the content of self-disclosure was relevant to the conversation because the content of work was statistically significantly more empathetic than the content of money in the self-disclosure attribute.

Finally, H4, ``The empathy capacity of the agent does not change with the attribute of self-disclosure, but it changes with the type of relationship presented," supported the hypothesis. 
The fact that this hypothesis was supported is a strength of this study. In this experiment, the relationship with the agent was only presented in short phrases via text message. 
Nevertheless, when participants perceived the agent as a cooperating partner, they indicated that they thought the agent had the capacity for empathy. 
In other words, the results showed that people evaluated the empathy capacity of the agent in advance on the basis of their relationship with the agent, simply by recognizing the agent's position. 
This indicated the possibility that the evaluation of the agent is determined in advance when the person and the agent actually interact.

%\subsection{Limitations}
A limitation of this study is that participants observed the agent's self-introduction by watching a video. 
Because the results of this study focused on changes in the situation before an actual task, participants did not perform an actual task with the agent. 
If a collaborative task had been added after this experiment, trust in the agent could have changed, and further changes in empathy toward the agent and the agent's ability to empathize with the participant could have occurred after the task. 
Future research will also focus on changes in trust and empathy after a task for both the cooperative and competitive tasks, and we will propose a design that allows agents used in society to develop appropriate relationships with people.

\section{CONCLUSIONS}
To establish appropriate trust and empathy relationships with anthropomorphic agents, including AI and robots, it is important for people to gain knowledge about the agents. 
An experiment was conducted with two inter-participant factors: the attribute of self-disclosure and the relationship with the agent. 
The number of levels for each factor was six levels of self-disclosure attributes and two levels of relationship with the agent. 
The dependent variables were trust and empathy toward the agent and the agent's capacity for empathy. 
The results showed that there were no significant differences in trust in the agent regardless of the attribute of self-disclosure or relationship with the agent. 
However, there was a main effect of the attribute of self-disclosure on empathy toward the agent. In addition, a main effect was found for the agent's ability to empathize with the participant regarding the relationship with the agent. 
These results supported our hypothesis. 
This study showed that when humans perceive the empathy capacity of the agent, simply being presented with the agent's relationship to the agent prior to the task changes their evaluation of the empathy capacity of the agent. 
This is an example of the importance of designing before the actual task to be performed when using agents in the future.

\section*{ACKNOWLEDGMENTS}
This work was partially supported by JST, CREST (JPMJCR21D4), Japan.
This work was also supported by JST, the establishment of university fellowships towards the creation of science technology innovation, Grant Number JPMJFS2136.

\addtolength{\textheight}{0cm}   % This command serves to balance the column lengths
                                  % on the last page of the document manually. It shortens
                                  % the textheight of the last page by a suitable amount.
                                  % This command does not take effect until the next page
                                  % so it should come on the page before the last. Make
                                  % sure that you do not shorten the textheight too much.

%%%%%%%%%%%%%%%%%%%%%%%%%%%%%%%%%%%%%%%%%%%%%%%%%%%%%%%%%%%%%%%%%%%%%%%%%%%%%%%%

%%%%%%%%%%%%%%%%%%%%%%%%%%%%%%%%%%%%%%%%%%%%%%%%%%%%%%%%%%%%%%%%%%%%%%%%%%%%%%%%

%%%%%%%%%%%%%%%%%%%%%%%%%%%%%%%%%%%%%%%%%%%%%%%%%%%%%%%%%%%%%%%%%%%%%%%%%%%%%%%%

%%%%%%%%%%%%%%%%%%%%%%%%%%%%%%%%%%%%%%%%%%%%%%%%%%%%%%%%%%%%%%%%%%%%%%%%%%%%%%%%

\bibliographystyle{IEEEtran}
\bibliography{test}

\begin{thebibliography}{10}
\providecommand{\url}[1]{#1}
\csname url@rmstyle\endcsname
\providecommand{\newblock}{\relax}
\providecommand{\bibinfo}[2]{#2}
\providecommand\BIBentrySTDinterwordspacing{\spaceskip=0pt\relax}
\providecommand\BIBentryALTinterwordstretchfactor{4}
\providecommand\BIBentryALTinterwordspacing{\spaceskip=\fontdimen2\font plus
\BIBentryALTinterwordstretchfactor\fontdimen3\font minus \fontdimen4\font\relax}
\providecommand\BIBforeignlanguage[2]{{%
\expandafter\ifx\csname l@#1\endcsname\relax
\typeout{** WARNING: IEEEtran.bst: No hyphenation pattern has been}%
\typeout{** loaded for the language `#1'. Using the pattern for}%
\typeout{** the default language instead.}%
\else
\language=\csname l@#1\endcsname
\fi
#2}}

\bibitem{Kaplan23}
\BIBentryALTinterwordspacing
A.~D. Kaplan, T.~T. Kessler, J.~C. Brill, and P.~A. Hancock, ``Trust in artificial intelligence: Meta-analytic findings,'' \emph{Human Factors}, vol.~65, no.~2, pp. 337--359, 2023, pMID: 34048287. [Online]. Available: \url{https://doi.org/10.1177/00187208211013988}
\BIBentrySTDinterwordspacing

\bibitem{Kähkönen21}
\BIBentryALTinterwordspacing
T.~Kähkönen, K.~Blomqvist, N.~Gillespie, and M.~Vanhala, ``Employee trust repair: A systematic review of 20 years of empirical research and future research directions,'' \emph{Journal of Business Research}, vol. 130, pp. 98--109, 2021. [Online]. Available: \url{https://www.sciencedirect.com/science/article/pii/S014829632100179X}
\BIBentrySTDinterwordspacing

\bibitem{Bunting21}
\BIBentryALTinterwordspacing
H.~Bunting, J.~Gaskell, and G.~Stoker, ``Trust, mistrust and distrust: A gendered perspective on meanings and measurements,'' \emph{Frontiers in Political Science}, vol.~3, 2021. [Online]. Available: \url{https://www.frontiersin.org/articles/10.3389/fpos.2021.642129}
\BIBentrySTDinterwordspacing

\bibitem{Nomura16}
T.~Nomura, T.~Kanda, H.~Kidokoro, Y.~Suehiro, and S.~Yamada, ``Why do children abuse robots?'' \emph{Interaction Studies}, vol.~17, no.~3, pp. 347--369, 2016.

\bibitem{Omdahl95}
B.~L. Omdahl, \emph{Cognitive appraisal, emotion, and empathy}, 1st~ed.\hskip 1em plus 0.5em minus 0.4em\relax New York: Psychology Press, 1995.

\bibitem{Preston02}
S.~D. Preston and F.~B.~M. de~Waal, ``Empathy: Its ultimate and proximate bases,'' \emph{Behavioral and Brain Sciences}, vol.~25, no.~1, p. 1–20, 2002.

\bibitem{Gillath21}
\BIBentryALTinterwordspacing
O.~Gillath, T.~Ai, M.~S. Branicky, S.~Keshmiri, R.~B. Davison, and R.~Spaulding, ``Attachment and trust in artificial intelligence,'' \emph{Computers in Human Behavior}, vol. 115, p. 106607, 2021. [Online]. Available: \url{https://www.sciencedirect.com/science/article/pii/S074756322030354X}
\BIBentrySTDinterwordspacing

\bibitem{Asan20}
\BIBentryALTinterwordspacing
O.~Asan, A.~E. Bayrak, and A.~Choudhury, ``Artificial intelligence and human trust in healthcare: Focus on clinicians,'' \emph{J Med Internet Res}, vol.~22, no.~6, p. e15154, Jun 2020. [Online]. Available: \url{https://doi.org/10.2196/15154}
\BIBentrySTDinterwordspacing

\bibitem{Maehigashi22-1}
A.~Maehigashi, T.~Tsumura, and S.~Yamada, ``Effects of beep-sound timings on trust dynamics in human-robot interaction,'' in \emph{Social Robotics}, F.~Cavallo, J.-J. Cabibihan, L.~Fiorini, A.~Sorrentino, H.~He, X.~Liu, Y.~Matsumoto, and S.~S. Ge, Eds.\hskip 1em plus 0.5em minus 0.4em\relax Cham: Springer Nature Switzerland, 2022, pp. 652--662.

\bibitem{Maehigashi22-2}
\BIBentryALTinterwordspacing
------, ``Experimental investigation of trust in anthropomorphic agents as task partners,'' in \emph{Proceedings of the 10th International Conference on Human-Agent Interaction}, ser. HAI '22.\hskip 1em plus 0.5em minus 0.4em\relax New York, NY, USA: Association for Computing Machinery, 2022, p. 302–305. [Online]. Available: \url{https://doi.org/10.1145/3527188.3563921}
\BIBentrySTDinterwordspacing

\bibitem{Maehigashi22-3}
A.~Maehigashi, ``The nature of trust in communication robots: Through comparison with trusts in other people and ai systems,'' in \emph{2022 17th ACM/IEEE International Conference on Human-Robot Interaction (HRI)}, 2022, pp. 900--903.

\bibitem{Okamura20-1}
K.~Okamura and S.~Yamada, ``Adaptive trust calibration for human-{AI} collaboration,'' \emph{PLOS ONE}, vol.~15, no.~2, pp. 1--20, 2020.

\bibitem{Okamura20-2}
------, ``Empirical evaluations of framework for adaptive trust calibration in human-ai cooperation,'' \emph{IEEE Access}, vol.~8, pp. 220\,335--220\,351, 2020.

\bibitem{Tsumura23-1}
\BIBentryALTinterwordspacing
T.~Tsumura and S.~Yamada, ``Influence of agent’s self-disclosure on human empathy,'' \emph{PLOS ONE}, vol.~18, no.~5, pp. 1--24, 05 2023. [Online]. Available: \url{https://doi.org/10.1371/journal.pone.0283955}
\BIBentrySTDinterwordspacing

\bibitem{Tsumura23-2}
------, ``Influence of anthropomorphic agent on human empathy through games,'' \emph{IEEE Access}, vol.~11, pp. 40\,412--40\,429, 2023.

\bibitem{Paiva17}
A.~Paiva, I.~Leite, H.~Boukricha, and I.~Wachsmuth, ``Empathy in virtual agents and robots: A survey,'' \emph{ACM Trans. Interact. Intell. Syst.}, vol.~7, no.~3, 2017.

\bibitem{Rahmanti22}
A.~R. Rahmanti, H.-C. Yang, B.~S. Bintoro, A.~A. Nursetyo, M.~S. Muhtar, S.~Syed-Abdul, and Y.-C.~J. Li, ``Slimme, a chatbot with artificial empathy for personal weight management: System design and finding,'' \emph{Frontiers in Nutrition}, vol.~9, 2022.

\bibitem{Jáuregui21}
D.~A. Gómez~Jáuregui, F.~Dollack, and M.~Perusquía-Hernández, ``Robot mirroring: Improving well-being by fostering empathy with an artificial agent representing the self,'' in \emph{2021 9th International Conference on Affective Computing and Intelligent Interaction Workshops and Demos (ACIIW)}, 2021, pp. 1--7.

\bibitem{Pan20}
W.~Pan, B.~Feng, V.~S. Wingate, and S.~Li, ``What to say when seeking support online: A comparison among different levels of self-disclosure,'' \emph{Frontiers in Psychology}, vol.~11, 2020.

\bibitem{Ruissen16}
M.~I. Ruissen and E.~R.~A. de~Bruijn, ``Competitive game play attenuates self-other integration during joint task performance,'' \emph{Frontiers in Psychology}, vol.~7, 2016.

\bibitem{Collins94}
N.~L. Collins and L.~C. Miller, ``\BIBforeignlanguage{en}{Self-disclosure and liking: a meta-analytic review},'' \emph{\BIBforeignlanguage{en}{Psychol Bull}}, vol. 116, no.~3, pp. 457--475, Nov. 1994.

\bibitem{Meng21}
\BIBentryALTinterwordspacing
J.~Meng and Y.~N. Dai, ``{Emotional Support from AI Chatbots: Should a Supportive Partner Self-Disclose or Not?}'' \emph{Journal of Computer-Mediated Communication}, vol.~26, no.~4, pp. 207--222, 05 2021. [Online]. Available: \url{https://doi.org/10.1093/jcmc/zmab005}
\BIBentrySTDinterwordspacing

\bibitem{Davis99}
R.~Davis, ``Web-based administration of a personality questionnaire: Comparison with traditional methods,'' \emph{Behavior Research Methods, Instruments, \& Computers}, vol.~31, pp. 572--577, 1999.

\bibitem{Jourard71}
S.~M. Jourard, \emph{Self-disclosure: An experimental analysis of the transparent self.}\hskip 1em plus 0.5em minus 0.4em\relax John Wiley, 1971.

\bibitem{Ullman19}
\BIBentryALTinterwordspacing
D.~Ullman and B.~F. Malle, ``Measuring gains and losses in human-robot trust: Evidence for differentiable components of trust,'' in \emph{2019 14th ACM/IEEE International Conference on Human-Robot Interaction (HRI)}, ser. 2019 14th ACM/IEEE International Conference on Human-Robot Interaction (HRI), 2019, pp. 618--619. [Online]. Available: \url{https://doi.org/10.1109/HRI.2019.8673154}
\BIBentrySTDinterwordspacing

\bibitem{Komiak06}
S.~Y.~X. Komiak and I.~Benbasat, ``The effects of personalizaion and familiarity on trust and adoption of recommendation agents,'' \emph{MIS Q.}, vol.~30, pp. 941--960, 2006.

\end{thebibliography}

\end{document}